\documentclass[11pt,letterpaper]{article}%
\usepackage{setspace}
\usepackage{color}
\usepackage{amsmath}
\usepackage{amsfonts}
\usepackage{subcaption}
\usepackage{comment}
\usepackage{booktabs}
\usepackage{verbatim}
\usepackage{amssymb}
\usepackage{graphicx}%
\providecommand{\U}[1]{\protect\rule{.1in}{.1in}}
\begin{document}

\title{Analytic crystals of solitons in the four dimensional \\gauged non-linear sigma model}
\author{Fabrizio Canfora$^{1}$, Seung Hun Oh$^{2}$, Aldo Vera$^{3}$\\$^{1}$\textit{Centro de Estudios Cient\'{\i}ficos (CECS), Casilla 1469,
Valdivia, Chile,}\\$^{2}$\textit{Institute of Convergence Fundamental Studies, School of Liberal
Arts,} \\\textit{Seoul National University of Science and Technology, Seoul 01811,
Korea,}\\$^{3}$\textit{Departamento de F\'{\i}sica, Universidad de Concepci\'{o}n,
Casilla 160-C, Concepci\'{o}n, Chile.}\\{\small canfora@cecs.cl, shoh.physics@gmail.com, aldovera@udec.cl}}
\maketitle

\begin{abstract}
The first analytic topologically non-trivial solutions in the
(3+1)-dimensional gauged non-linear sigma model representing multi-solitons at
finite volume with manifest ordered structures generating their own
electromagnetic field are presented. The complete set of seven coupled
non-linear field equations of the gauged non-linear sigma model together with
the corresponding Maxwell equations are reduced in a self-consistent way to
just one linear Schrodinger-like equation in two dimensions. The corresponding
two dimensional periodic potential can be computed explicitly in terms of the
solitons profile. The present construction keeps alive the topological charge
of the gauged solitons. Both the energy density and the topological charge
density are periodic and the positions of their peaks show a crystalline
order. These solitons describe configurations in which (most of) the
topological charge and total energy are concentrated within three-dimensional
tube-shaped regions. The electric and magnetic fields vanish in the center of
the tubes and take their maximum values on their surface while the
electromagnetic current is contained within these tube-shaped regions.
Electromagnetic perturbations of these families of gauged solitons are shortly discussed.

\newpage

\end{abstract}
\tableofcontents

\section{Introduction}

One of the most relevant field theories (both due to its predictive power as
well as its non-trivial topological features) is the non-linear sigma model.
It is a very effective tool from high energy physics to statistical mechanics
systems like quantum magnetism, the quantum hall effect, meson interactions,
superfluid $^{3}$He and string theory (see \cite{manton} \cite{BaMa}). It was
introduced in particle physics to describe the low-energy dynamics of pions
(see for instance \cite{2}, \cite{3} and references therein). On the other
hand, as Skyrme noticed \cite{skyrme}, such a theory does not possess
solitonic solutions in flat, topologically trivial (3+1)-dimensional
spacetimes. This can be shown using the Derrick's scaling argument \cite{4}
(although Skyrme understood this before Derrick himself). That's why Skyrme
introduced his famous Skyrme term \cite{skyrme}. However, it is worth
emphasizing that the beautiful currents-algebraic arguments by Witten
\cite{witten0} (see also \cite{bala0}, \cite{Bala1}, \cite{ANW} and references
therein)\ to show that the solitons of this theory should be quantized at a
semi-classical level as Fermions, and that such a theory describe the
low-energy limit of QCD does not make explicit use of the Skyrme term itself but
only of the fact that stable solitons with non-trivial third homotopy class
exist (together with the well known Wess-Zumino-Witten term): a detailed
review of these issues can be found in \cite{BaMa}. It is thus of great
theoretical interest to search for a natural mechanism to avoid the Derrick
scaling argument within the non-linear sigma model.

A topic which has a great theoretical as well as phenomenological importance
is (a proper theoretical understanding of) crystals of solitons (which are
very useful, for instance, in the description of cold and dense nuclear matter
as a function of the topological charge \cite{R1}, \cite{R2}). That's why it
is relevant to analyze configurations with many solitons coexisting within a
fixed spatial volume. In this very interesting but extremely complicated phase
the most common analytic methods are basically useless and so solitons living
within a finite volume in (3+1)-dimensions with a finite topological charge
are usually studied numerically. The importance of this issue comes from the
fact that crystals of solitons are expected to appear when, within a fixed
spatial volume, the topological charge is high enough. Until very recently,
such solitonic crystals could only be found in two-dimensions (see
\cite{toy1}, \cite{toy2}\ and references therein). Powerful approximation
schemes developed in \cite{aprox0}, \cite{aprox1}, \cite{aprox2},
\cite{aprox3}, \cite{aprox4}, \cite{aprox5}, \cite{aprox5.1}, \cite{aprox5.2},
shed light on many properties of these crystals but analytic examples are very rare.

It is well known that, as the density increases, quite complicated ordered
structures denoted as \textit{nuclear pasta phases} appear as well (see
\cite{pasta1}, \cite{pasta2}, \cite{pasta3}, \cite{pasta4} and references
therein). The numerical works on these structures reveal the presence of
\textquotedblleft baryonic tubes\textquotedblright\ (\textit{nuclear
spaghetti}) in which most of the baryonic charge is concentrated within
tube-shaped regions in three dimensions\footnote{\textit{Nuclear lasagna} and
\textit{nuclear gnocchi} phases are also known to appear: see the references
quoted above.}. These configurations with high topological charge living
within finite volumes are also very difficult to describe analytically (the
only available results in these phases have been obtained numerically).

Needless to say, even less is known when the electromagnetic interactions are
taken into account. It would be very helpful to have some analytic control on
crystals of \textit{gauged solitons} in (3+1) dimensions, namely, exact
solutions of theory with topological charge which are minimally coupled to
$U(1)$ gauge theories and with ordered structure. In particular, it would be
quite an achievement to construct analytic examples of gauged solitons with
high topological charge and, at the same time, with crystalline order. This
would shed new light on the electromagnetic properties of strongly interacting
solitons in a self-consistent way. Very few exact results are known in the
literature: these results are derived either in 2+1 dimensions or (in higher
dimensions) when some extra symmetries (such as SUSY) are available (see
\cite{gaugsol0}, \cite{gaugsol}, \cite{gaugsol1}, \cite{gaugsol2},
\cite{gaugsol3}, \cite{gaugsol4}, \cite{gaugsol5}, \cite{gaugsol6},
\cite{gaugsol7}, \cite{gaugsol8} and references therein). As one can see from
the above references, unless suitable BPS bounds which can be saturated are
available, gauged solitons can only be constructed numerically. Thus, in
particular, there is no analytic example of a crystal of gauged solitons in
3+1 dimensions in the $SU(2)$ non-linear sigma model minimally coupled to a
$U(1)$ gauge field (which will be denoted in the following as gauged
non-linear sigma model). This topic is not just of academic interest since, in
many applications (from plasma physics to nuclear physics and astrophysics),
the analysis of the electromagnetic properties of strongly interacting
solitons is extremely important and very challenging even from the numerical
point of view so that analytic methods able to simplify such systems are welcome.

As we will show in the following sections, one can find analytically crystal
of gauged solitons even when no \textquotedblleft saturable" BPS bound is
available if the Derrick argument is avoided in a suitable way. There are
three physically meaningful approaches to avoid Derrick's argument in the
non-linear sigma model which will be combined together in the following sections.

The first one is to search for a time-periodic ansatz such that the energy
density of the configuration is still static, as it happens for boson stars
\cite{5} (in the simpler case of $U(1)$-charged scalar field see \cite{6} and
references therein).

The second approach corresponds to consider the minimal coupling of the
non-linear sigma model with a $U(1)$ gauge field describing the
electromagnetic properties of the low energy limit of QCD.

The third one is to analyze the model at a finite fixed spatial volume.
Indeed, non-trivial boundary conditions at finite volume can also break the
Derrick scaling argument.

These three ways to avoid the Derrick no-go argument can be combined using the
generalized hedgehog ansatz introduced in \cite{canfora2}, \cite{56},
\cite{56b}, \cite{58}, \cite{58b}, \cite{ACZ}, \cite{CanTalSk1},
\cite{canfora10}, \cite{Giacomini:2017xno}, \cite{ACLV}, \cite{Fab1},
\cite{gaugsk}, \cite{Canfora:2018clt} \cite{lastEPJC}. This approach allowed
the construction of analytical solutions that describe multi-solitons at
finite density for both, in the non-linear sigma model as well as in the
Skyrme model (coupled to the Maxwell theory in \cite{gaugsk} and
\cite{Canfora:2018clt}). Even more, recently, configurations of analytic
multi-Skyrmions with crystalline order were constructed in \cite{crystal1}.

In this work, following the steps of \cite{gaugsk}, \cite{Canfora:2018clt} and
\cite{crystal1}, we will construct, for the non-linear sigma model, analytic
gauged multi-soliton solutions at finite density with crystalline structure.
Using the generalized hedgehog ansatz in a sector with non-vanishing
topological charge, the complete set of seven coupled field equations are
reduced in a self-consistent way to one linear Schrodinger-like equation with
an effective two dimensional periodic potential which can be computed
explicitly in terms of the solitons profile. These gauged solitons describe
configurations in which (most of) the topological charge and total energy are
concentrated within tube-shaped regions whose positions are regular in space
(very much like -a charged version of- the nuclear spaghetti in \cite{pasta1},
\cite{pasta2}, \cite{pasta3}, \cite{pasta4}). On the other hand, the electric
and magnetic fields vanish in the center of these tubes and take their maximum
values on their surface while the electromagnetic current is contained within
these tube-shaped regions. The present analytic construction of ordered arrays
of gauged tubes allows, in particular, the analysis of electromagnetic
perturbations of these crystals. It will be shown that the Maxwell field
perceives these gauged solitons as an effective periodic medium whose
properties can be studied explicitly.

This paper is organized as follows: In the next section the gauged non-linear
sigma model is introduced. In the third section, using the generalized
hedgehog ansatz, we reduce the equations system to an only one linear
Schrodinger-like equation with an effective potential. Also we compute the
energy and the topological density. In section IV, through some plots of
particular configurations, we clarify the physical meaning of the solutions
here constructed and we show how a crystalline structure of multi-solitons
emerge. We also analyze the perturbative stability of these solutions. In the
final section some conclusions will be drawn.

\section{The gauged non-linear sigma model}

A very important question is whether or not it is possible to construct
consistently crystals of gauged solitons and, at the same time, the
corresponding electromagnetic fields in the gauged non-linear sigma model. The
importance in many phenomenologically relevant situations to analyze the
interactions between solitons of the low-energy limit of QCD with $U(1)$ gauge
fields makes mandatory the task to arrive at a deeper understanding of the
gauged non-linear sigma model (classic references are \cite{witten0},
\cite{Witten}, \cite{gipson}, \cite{goldstone}, \cite{dhoker} and
\cite{rubakov}).

Until very recently mainly numerical tools were employed (see \cite{gaugesky1}%
, \cite{gaugesky2}\ and references therein) to analyze these configurations.
Here we will show that it is possible to construct crystals of gauged solitons
in a finite volume using a time-dependent ansatz for the $U$ field that allows
to have non-vanishing topological charge and, at the same time, leads to a
static energy-momentum tensor as well as to static electric and magnetic fields.

\vspace{.2cm}

The action of the $U(1)$ gauged non-linear sigma model in four dimensions is
\begin{align}
S  &  =\int d^{4}x\sqrt{-g}\left[  \frac{K}{4}\mathrm{Tr}\left(  L^{\mu}%
L_{\mu}\right)  - m^{2}\mathrm{Tr}\left(  U+U^{-1} \right)  -\frac{1}{4}%
F_{\mu\nu}F^{\mu\nu}\right]  \ ,\label{sky1}\\
L_{\mu}  &  =U^{-1}D_{\mu}U\ ,\ \ \ D_{\mu}=\nabla_{\mu}+A_{\mu}\left[
t_{3},\ .\ \right]  \ ,\label{sky2}\\
U  &  \in SU(2)\ ,\ \ L_{\mu}=L_{\mu}^{j}t_{j}\ ,\ \ t_{j}=i\sigma
_{j}\ ,\ F_{\mu\nu}=\partial_{\mu}A_{\nu}-\partial_{\nu}A_{\mu}\ ,
\label{sky2.5}%
\end{align}
where $g$ is the metric determinant, $m$ is the Pions mass,
%$\mathbf{1}_{2}$ is the $2\times2$ identity matrix,
$A_{\mu}$ is the gauge potential, $\nabla_{\mu}$ is the partial derivative and
$\sigma_{i}$ are the Pauli matrices.

The energy-momentum tensor is given by
\begin{equation}
T_{\mu\nu}=-\frac{K}{2}\mathrm{Tr}\left[  L_{\mu}L_{\nu}-\frac{1}{2}g_{\mu\nu
}L^{\alpha}L_{\alpha}\right]  -m^{2} \mathrm{Tr}\biggl[ g_{\mu\nu}(U+U^{-1})
\biggl] +\bar{T}_{\mu\nu}\ ,\nonumber
\end{equation}
with
\begin{equation}
\bar{T}_{\mu\nu}=F_{\mu\alpha}F_{\nu}^{\;\alpha}-\frac{1}{4}F_{\alpha\beta
}F^{\alpha\beta}g_{\mu\nu}\ , \label{tmunu(1)}%
\end{equation}
being the electromagnetic energy-momentum tensor.

The field equations read
\begin{equation}
D_{\mu}L^{\mu}+\frac{2 m^{2}}{K} \left(  U-U^{-1}\right)  =0 \ , \label{NLSM}%
\end{equation}
\begin{equation}
\nabla_{\mu}F^{\mu\nu}=J^{\nu}\ , \label{maxwellNLSM}%
\end{equation}
where the current $J^{\mu}$ is given by
\begin{align}
J^{\mu}=\frac{K}{2}\text{Tr}\left[  \widehat{O}L^{\mu}\right]  \ ,\qquad
\widehat{O}=U^{-1}t_{3}U-t_{3}\ . \label{current}%
\end{align}

It is worth to note that when the gauge potential reduces to a constant along
the time-like direction, the field equations in Eqs. (\ref{NLSM}) and
(\ref{maxwellNLSM}) describe the non-linear sigma model at a finite isospin
chemical potential.

As usual in the literature, the term \textit{gauged solitons} will refer to
smooth regular solutions of the coupled system in Eqs. (\ref{NLSM}) and
(\ref{maxwellNLSM}) possessing a non-vanishing baryon charge (defined below in
Eq. (\ref{new4.1})). We will see later that, for the class of solutions
presented in this work, the number of peaks in the energy density is related
with the topological charge itself. The gauged solitons are considered to be
static if the energy density (and, more generically, the energy-momentum
tensor) does not depend on time. In other words,\textit{ a gauged soliton is
considered to be static if it corresponds to a static distribution of
energy-density}.

%\subsection{Gauged topological charge}
\vspace{.2cm} The correct expression for the topological charge of the gauged
non-linear sigma model has been constructed in \cite{Witten} (see also the
pedagogical analysis in \cite{gaugesky1}):%
\begin{equation}
W=\frac{1}{24\pi^{2}}\int_{\Sigma}\rho_{B}\ , \label{new4.1}%
\end{equation}
where
\begin{equation}
\rho_{B}=\epsilon^{ijk}\text{Tr}\biggl[\left(  U^{-1}\partial_{i}U\right)
\left(  U^{-1}\partial_{j}U\right)  \left(  U^{-1}\partial_{k}U\right)
-\partial_{i}\left[  3A_{j}t_{3}\left(  U^{-1}\partial_{k}U+\left(
\partial_{k}U\right)  U^{-1}\right)  \right]  \biggl]\ . \label{new4.1.1}%
\end{equation}

Note that the second term in Eq. (\ref{new4.1.1}) guarantees both the
conservation and the gauge invariance of the topological charge. When $\Sigma$
is space-like, $W$ is the baryon charge of the configuration.

\section{Gauged crystals}

\subsection{The ansatz}

The main physical motivation of the present work is to study crystals of
gauged solitons living within a finite volume. The easiest way to take into
account finite volume effects is to use the flat metric defined below:
\begin{equation}
ds^{2}=-dt^{2}+L^{2}\left(  dr^{2}+d\theta^{2}+d\phi^{2}\right)  \ ,
\label{Minkowski}%
\end{equation}
where $4\pi^{3}L^{3}$ is the volume of the box in which the gauged solitons
are living. The adimensional coordinates $r$, $\theta$ and $\phi$ have the
ranges
\begin{equation}
0\leq r\leq2\pi\ ,\quad0\leq\theta\leq\pi\ ,\quad0\leq\phi\leq2\pi\ .
\label{period0}%
\end{equation}
Following the strategy of \cite{Fab1}, \cite{gaugsk} and
\cite{Canfora:2018clt}, the boundary conditions in the $\theta$ direction will
be chosen to be Dirichlet while in the $r$ and $\phi$\ directions they can be
both periodic and anti-periodic.

Combining the strategy of \cite{Fab1}, \cite{gaugsk}, \cite{Canfora:2018clt}
with \cite{crystal1}, one arrives at the following ansatz for the gauged
crystals:%
\begin{align}
U^{\pm1}(x^{\mu})  &  =\cos\left(  \alpha\right)  \mathbf{1}_{2}\pm\sin\left(
\alpha\right)  n^{i}t_{i}\ ,\ \ n^{i}n_{i}=1\ ,\label{standard1}\\
n^{1}  &  =\sin\Theta\cos\Phi\ ,\ \ n^{2}=\sin\Theta\sin\Phi\ ,\ \ n^{3}%
=\cos\Theta\ , \label{standard1.1}%
\end{align}%
\begin{equation}
\alpha=\alpha\left(  r\right)  \ ,\ \Theta=q\theta\ ,\ \Phi=p\left(  \frac
{t}{L}-\phi\right)  \ ,\ q=2v+1\ ,\quad p,v\in\mathbb{N}\ ,\ p\neq0\ ,
\label{ans1}%
\end{equation}%
\begin{equation}
A_{\mu}=(u(r,\theta),0,0,-Lu(r,\theta))\ . \label{gaugcrystal2}%
\end{equation}
The periodic time dependence\footnote{With the above choice of the ansatz the
$U$ field is periodic in time since $U$ depends on time through $\sin\Phi$ and
$\cos\Phi$.} in Eq. (\ref{ans1}) deserves some comments. First of all, it is a
key technical assumption which allows to solve the field equations
analytically, as we will detail later. Secondly, the Derrick's famous no-go
theorem on the existence of solitons in non-linear scalar field theories is
avoided using a time-periodic ansatz such that the energy density of the
configuration is still static (see \cite{5}, \cite{6} and references therein).
The present ansatz defined in Eqs. (\ref{standard1}), (\ref{standard1.1}) and
(\ref{ans1}) has exactly this property (a direct computation shows that the
energy density does not depend on time). Thirdly, unlike what happens for the
usual bosons star ansatz for $U(1)$-charged scalar fields, the present ansatz
for $SU(2)$-valued scalar field also possesses a non-trivial topological charge.

\subsection{Field equations}

Quite remarkably, with the above ansatz for the gauge potential $A_{\mu}$ and
for the $SU(2)$-valued scalar field, it is possible to reduce the gauged
non-linear sigma model field equation in Eq. (\ref{NLSM}) to a single
integrable equation for the profile $\alpha$ as well as the Maxwell equations
in Eq. (\ref{maxwellNLSM}) to only one linear Schrodinger-like equation in
which the effective two-dimensional periodic potential can be computed
explicitly in terms of the profile $\alpha$ itself. More concretely, this
problem can be divided into two steps.

\textit{First step}: with the ansatz for the $SU(2)$-valued scalar field in
Eqs. (\ref{standard1}), (\ref{standard1.1}) and (\ref{ans1}) and for the gauge
potential in Eq. (\ref{gaugcrystal2}), the three gauged non-linear sigma model
field equations in Eq. (\ref{NLSM}) reduce to a single equation for the
profile $\alpha$, namely
\begin{equation}
\alpha^{\prime\prime}-\frac{q^{2}}{2}\sin(2\alpha)+\frac{4m^{2}}{K}\sin
(\alpha)=0\ .\label{sg1}%
\end{equation}
One may wonder why the $U(1)$ gauge potential does not enter explicitly in the
field equations in Eq. (\ref{NLSM}). This happens for the judicious choice of
the ansatz in Eqs. (\ref{ans1}) and (\ref{gaugcrystal2}). In particular, very
important relations which allow to decoupled Eq. (\ref{NLSM}) from Eq.
(\ref{maxwellNLSM}) are
\begin{align*}
A_{\mu}A^{\mu}  & =0\ ,\qquad\left(  \nabla_{\mu}\Phi\right)  \left(
\nabla^{\mu}\Phi\right)  =0\ ,\qquad A_{\mu}\nabla^{\mu}\Phi=0\ ,\\
\left(  \nabla_{\mu}\Theta\right)  \left(  \nabla^{\mu}\Phi\right)    &
=0\ ,\qquad  A_{\mu}\nabla^{\mu}\Theta=0 \ ,\qquad  A_{\mu}\nabla^{\mu}\alpha=0\ .
\end{align*}

Thanks to the above relations, one can first solve the equation of the gauged
non-linear sigma model explicitly. Then, once the $SU(2)$ valued scalar field
is known, the Maxwell equations reduce to a linear equation in which the
soliton plays the role of an effective potential. Note however, that \textit{
here no approximation has been made}: namely we are dealing with the complete
set of seven coupled non-linear field equations in Eqs. (\ref{NLSM}) and
(\ref{maxwellNLSM}) in which both the backreaction of the solitons on Maxwell
field and vice versa are explicitly taken into account in a self-consistent
way\footnote{Indeed, a direct computation shows that Eqs. (\ref{NLSM}) and
(\ref{maxwellNLSM}) with the ansatz in Eqs. (\ref{ans1}) and
(\ref{gaugcrystal2}) reduce exactly to Eqs. (\ref{sg1}) and
(\ref{sesseanewmax}). More details can be found in the Appendix.}.

Eq. (\ref{sg1}) can be integrated easily in terms of inverse elliptic
functions observing that it is equivalent to the following first order
equation
\begin{align}
\alpha^{\prime} &  =\pm\left[  2\left(  E_{0}-\frac{q^{2}}{4}\cos\left(
2\alpha\right)  +\frac{4m^{2}}{K}\cos(\alpha)\right)  \right]  ^{1/2}%
\ ,\label{sg2}\\
E_{0} &  >\frac{q^{2}}{4}+\frac{4m^{2}}{K}\Rightarrow\ \alpha^{\prime}%
\neq0\ \forall r\ .\label{sg2.1}%
\end{align}
The integration constant $E_{0}$ will be determined in the following
subsection (see Eq. (\ref{intconstcond}) below) requiring to have a
non-vanishing topological charge within a finite volume\footnote{It is worth
to note that the solutions of Eq. (\ref{sg2}) can be expressed in terms of
inverse elliptic functions (Jacobi functions). However, for the following
results we will not need these explicit expressions.}.

\textit{Second step}: it is perhaps even more surprising that the four Maxwell
equations in Eq. (\ref{maxwellNLSM}) with the current in Eq. (\ref{current})
corresponding to the $SU(2)$-valued field constructed in the previous step
reduce consistently to only one equation for $u(r,\theta)$ (defined in Eq.
(\ref{gaugcrystal2})):
\begin{equation}
\left(  \frac{\partial^{2}}{\partial r^{2}}+\frac{\partial^{2}}{\partial
\theta^{2}}\right)  u+Vu=\sigma\ , \label{max1}%
\end{equation}
where%
\[
V=\frac{2L}{p}\sigma\ ,\qquad\sigma=2pLK\sin^{2}(\alpha)\sin^{2}\left(
q\theta\right)  \ .
\]
It is useful to rewrite Eq. (\ref{max1}) as
\begin{equation}
\left(  \frac{\partial^{2}}{\partial r^{2}}+\frac{\partial^{2}}{\partial
\theta^{2}}\right)  \Psi+V\Psi=0 \ , \label{sesseanewmax}%
\end{equation}
where
\[
\Psi=\frac{2L}{p}u-1\ .
\]
The conclusion of this section is that the coupled system of seven field
equations made by the three field equations of the gauged non-linear sigma
model and the four Maxwell equations with the current corresponding to the
non-linear sigma model itself (in a sector with non-trivial topological
charge, see the next subsections) reduce exactly to just one linear equation
which can be regarded as a Schrodinger equation with a two-dimensional
periodic potential. The effective potential in Eq. (\ref{sesseanewmax}) is
known explicitly since the gauged non-linear sigma model equations are solved
by Eq. (\ref{sg2}). The physical interpretation of these result will be
discussed in the next section.

\subsection{Topological charge and energy density}

Taking into account Eqs. (\ref{standard1}), (\ref{standard1.1}), (\ref{ans1}),
(\ref{gaugcrystal2}) and (\ref{sg2}) the topological density defined in Eqs.
(\ref{new4.1}) and (\ref{new4.1.1}), for the configurations here constructed,
read
\begin{align}
\rho_{B}=\rho_{B}^{\text{NLSM}}+\rho_{B}^{\text{Maxwell}}\ , \label{bc0}
\end{align}
where the contributions of the non-linear sigma model and the Maxwell theory
are, respectively
\begin{align*}
\rho_{B}^{\text{NLSM}} &  =-12pq\sin(q\theta)\sin^{2}(\alpha)\partial
_{r}\alpha\ ,\\
\rho_{B}^{\text{Maxwell}} &  =12L\biggl[\biggl(2q\sin(q\theta)\sin^{2}%
(\alpha)u-\cos(q\theta)\partial_{\theta}u\biggl)\partial_{r}\alpha
-q\sin(\alpha)\cos(\alpha)\sin(q\theta)\partial_{r}u\biggl]\ ,
\end{align*}
which can be also written as
\[
\rho_{B}=3q\ \frac{\partial}{\partial r}\Big(p\sin(q\theta)\ \big(\sin
(2\alpha)-2\alpha\big)-2L\sin(q\theta)u\sin(2\alpha)\Big)-\frac{\partial
}{\partial\theta}\big(12L\alpha^{\prime}u\cos(q\theta)\big)\ .
\]
Thus, we can read the boundary conditions for the fields:
\begin{equation}
\alpha\left(  2\pi\right)  -\alpha\left(  0\right)  =n\pi\ ,\label{bc1}%
\end{equation}
and with this the topological charge becomes
\[
W=-np\times\Big(\frac{1-(-1)^{q}}{2}\Big)-\frac{L}{\pi}\int_{0}^{2\pi
}dr\ \alpha^{\prime}\Big((-1)^{q}\ u(r,\pi)-u(r,0)\Big)\ .
\]
If we assume a boundary condition for $u$ given by
\begin{equation}
u(r,\pi)=(-1)^{q}\ u(r,0)\ ,\label{bc2}%
\end{equation}
we obtain
\begin{equation}
W=%
\begin{cases}
-np & \text{if}\quad q\in2\mathbb{Z}+1\ ,\\[1.5ex]%
0 & \text{if}\quad q\in2\mathbb{Z}\ .
\end{cases}
\end{equation}
Now we note that, according to Eqs. (\ref{sg2}), (\ref{sg2.1}) and
(\ref{bc1}), the integration constant $E_{0}$ is fixed in terms of $n$ through
the relation
\begin{equation}
n\int_{0}^{\pi}\frac{1}{\eta(n\alpha,E_{0})}d\alpha=2\pi\ ,\quad\eta
(\alpha,E_{0})=\pm\biggl[2\left(  E_{0}-\frac{q^{2}}{4}\cos\left(
2\alpha\right)  +\frac{4m^{2}}{K}\cos(\alpha)\right)  \biggl]^{1/2}%
\ .\label{intconstcond}%
\end{equation}

On the other hand, the energy density $T_{00}$ reads%
\begin{equation}
T_{00}=T_{00}^{\text{NLSM}}+T_{00}^{\text{Maxwell}}\ , \label{t001}%
\end{equation}
where $T_{00}^{\text{NLSM}}$, $T_{00}^{\text{Maxwell}}$ are the energy density
of the gauged non-linear sigma model and the energy density contribution of
the Maxwell theory, respectively, which are given by
\begin{align}
T_{00}^{\text{NLSM}}\  &  =\frac{K}{2L^{2}}\biggl[\alpha^{\prime2}+2\sin
^{2}(\alpha)\sin^{2}(q\theta)(p-2Lu)^{2}+q^{2}\sin^{2}(\alpha) \biggl] + 4
m^{2} \cos(\alpha) \ ,\label{t002}\\
T_{00}^{\text{Maxwell}}\  &  =\frac{1}{L^{2}}\Big[\left(  \frac{\partial
u}{\partial r}\right)  ^{2}+\left(  \frac{\partial u}{\partial\theta}\right)
^{2}\Big]\ . \label{t003}%
\end{align}

In the next section, the plots of the energy density will clarify the physical
interpretation of the present multi-solitonic configurations.

\section{Gauged baryonic tubes}

In this section the physical interpretation of the gauged solitons is analyzed.

\subsection{Current, electric field and magnetic field}

The explicit expression of the topological density $\rho_{B}$ (in Eq.
(\ref{bc0})), the energy density $T_{00}$ (in Eqs. (\ref{t001}), (\ref{t002})
and (\ref{t003})) and the current $J_{\mu}$ (defined in Eq. (\ref{current}))
allows a clear physical interpretation of the above gauged solitons.
%\begin{align*}
%J^{\mu}=\frac{K}{2}\text{Tr}\left[  \widehat{O}L^{\mu}\right]  \ .
%\end{align*}
%Replacing the explicit solutions in Eqs. (\ref{standard1}),
%(\ref{standard1.1}), (\ref{gaugcrystal1}), (\ref{sg2}) and (\ref{gaugcrystal2}%
%) we get \textbf{XX please put XXX}%

In our configurations the non-vanishing components of the current are
\begin{align*}
J_{t}  &  =\frac{2K}{L}\sin^{2}(\alpha)\sin^{2}(q\theta)\left(  p-2Lu\right)
\ ,\\
J_{\phi}  &  =-2K\sin^{2}(\alpha)\sin^{2}(q\theta)\left(  p-2Lu\right)  \ .
\end{align*}
One can see both from the above expressions for energy density $T_{00}$, the
topological density $\rho_{B}$ and the current ($J_{t}$ and $J_{\phi}$) that
they are constant in the $\phi$ direction while they depend non-trivially on
the two spatial coordinates $r$ and $\theta$. For instance, the baryon density
$\rho_{B}$ is periodic in $r$ and $\theta$ while it is constant in the $\phi
$-direction. The positions of the maxima of $\rho_{B}$ are a regular
two-dimensional lattice (the same is true for $T_{00}$, $J_{t}$ and $J_{\phi}%
$). If we would include a three-dimensional contour plot of $\rho_{B}$ in the
present work\footnote{Unfortunately, we have been unable to reduce the size of
the three-dimensional contour plots below the 30 MB.}, it would reveal that
the regions of maximal $\rho_{B}$ are three-dimensional tubes of length $2\pi
L$ parallel to the $\phi$ direction (the same is true for $T_{00}$, $J_{t}$
and $J_{\phi}$). However, two-dimensional contour plots are quite enough to
understand the distribution of energy, topological density and currents of the
present (ordered arrays of) gauged tube-shaped solitons. The similarity with
the plots obtained numerically in the analysis of nuclear spaghetti phase is
quite amusing \cite{pasta1}, \cite{pasta2}, \cite{pasta3}, \cite{pasta4}. To
the best of authors knowledge, this is the first time that gauged solitonic
tubes are constructed analytically in (3+1) dimensions.

\begin{figure}[th]
\centering
\includegraphics[scale=0.25]{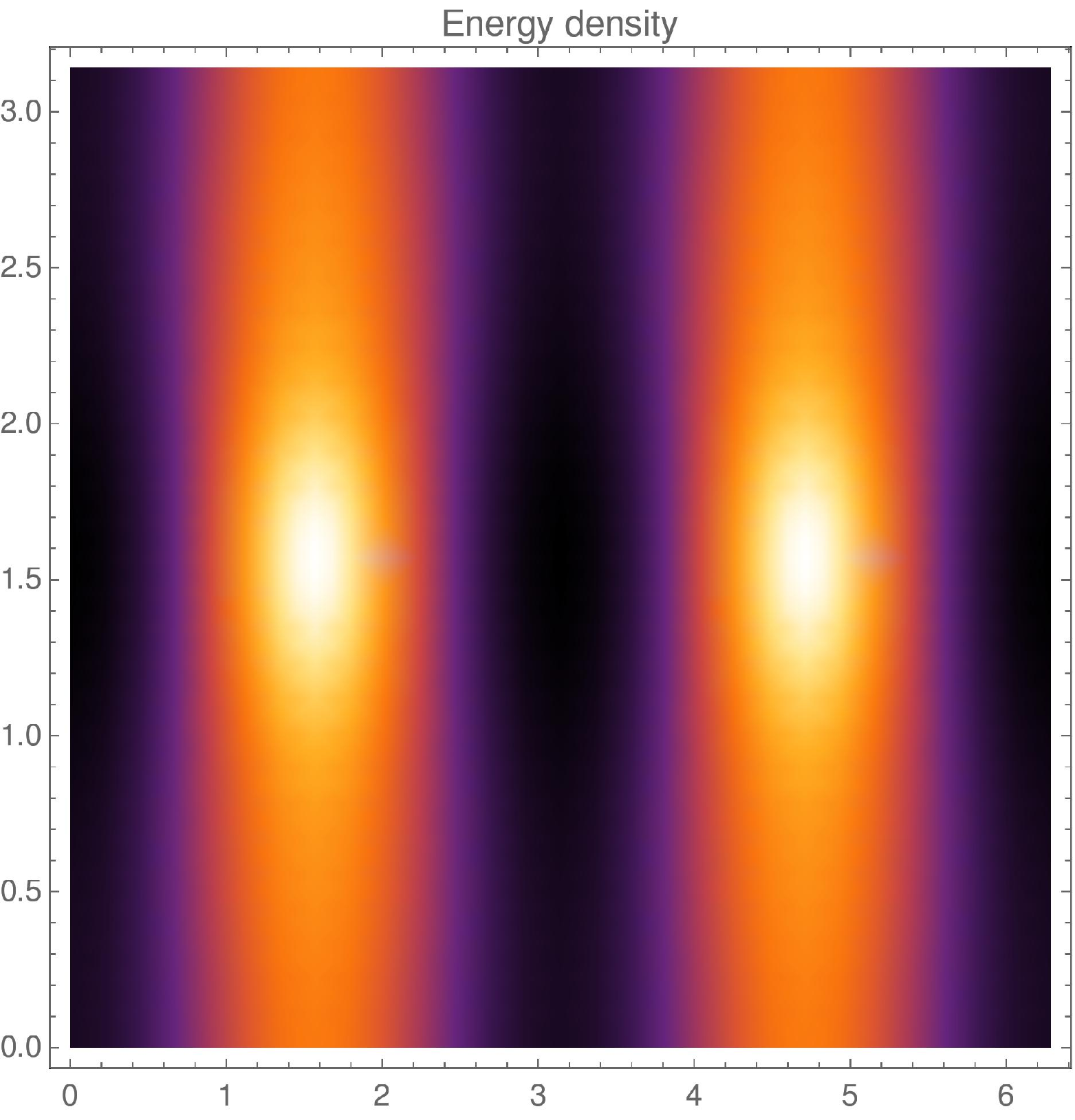}\quad\includegraphics[scale=0.35]{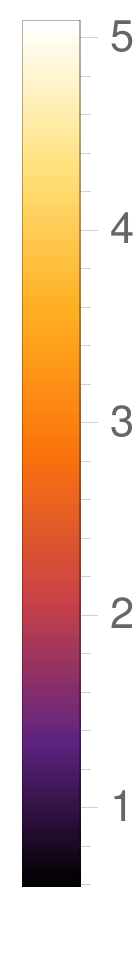}\quad
\includegraphics[scale=0.25]{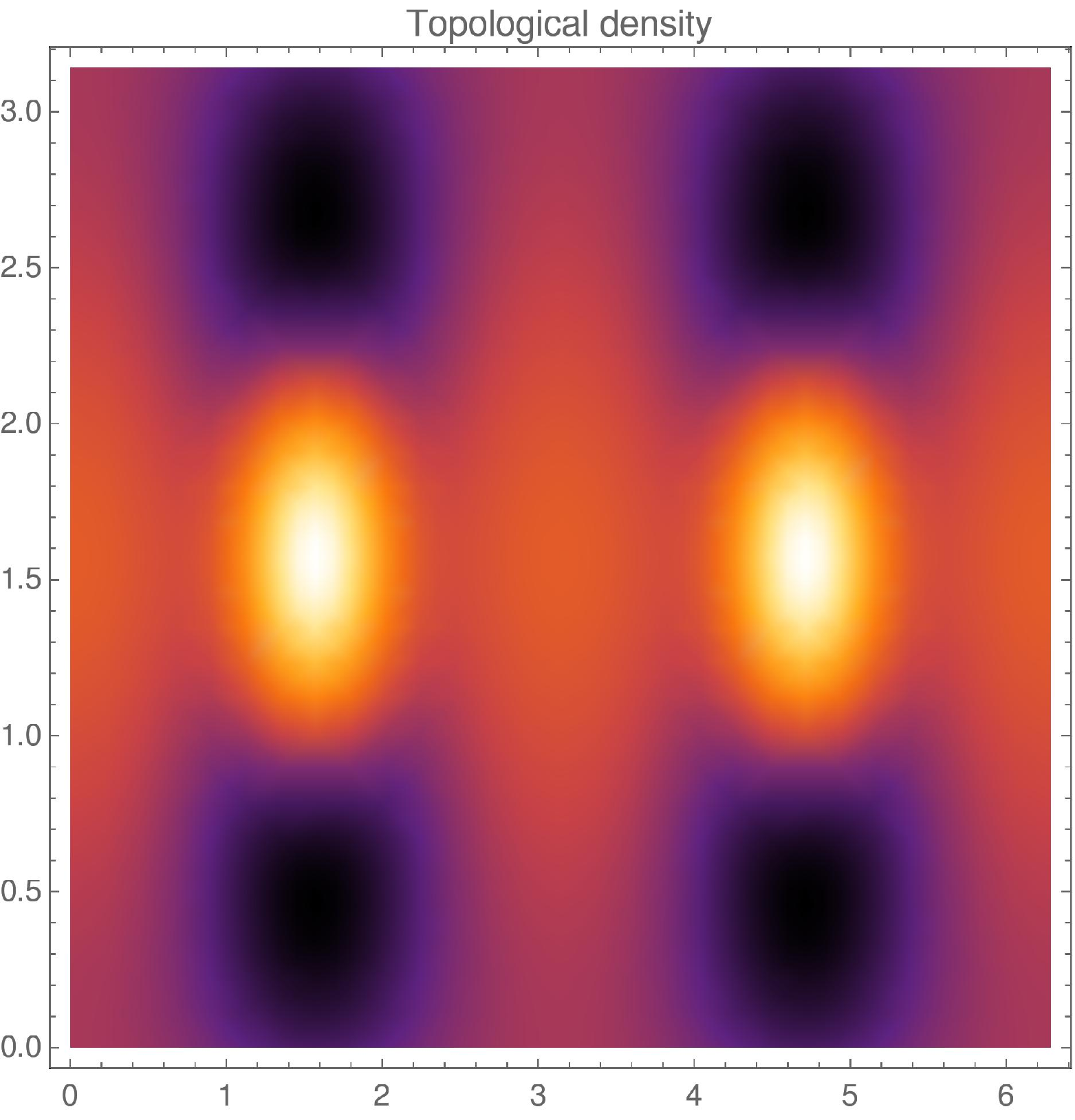}\quad\includegraphics[scale=0.35]{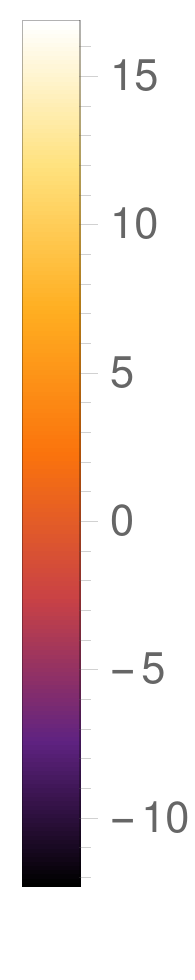}
\caption{Energy density and topological charge density of two gauged solitons,
with $n=2$, $m=0$, and $p=q=1$. Here we can see the periodic ordering of the
tube-shaped configurations.}%
\label{Ds}%
\end{figure}

\begin{figure}[th]
\centering
\includegraphics[scale=0.25]{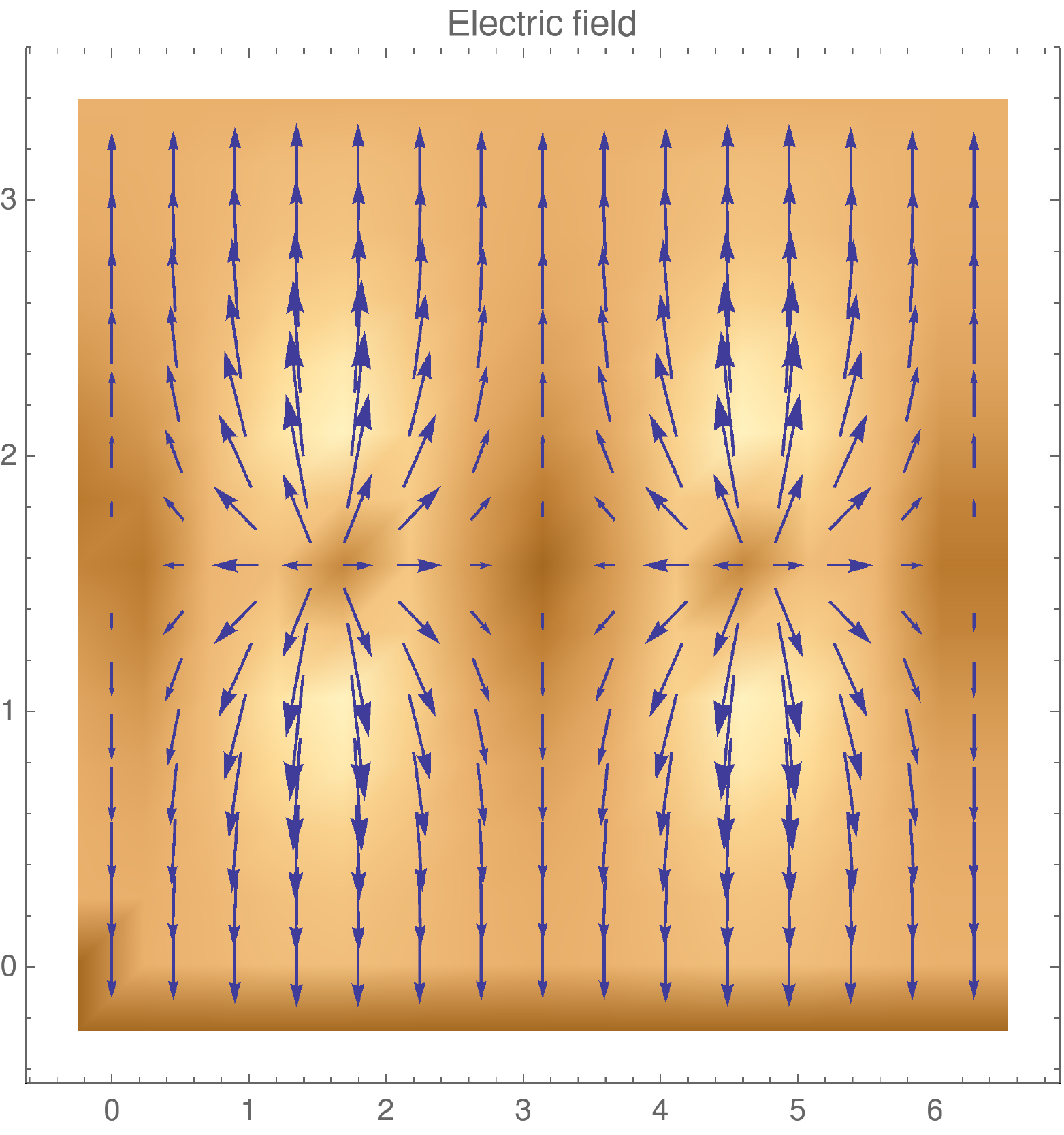}\quad
\includegraphics[scale=0.35]{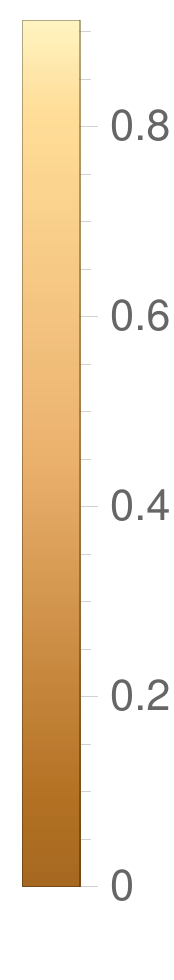}\quad
\includegraphics[scale=0.25]{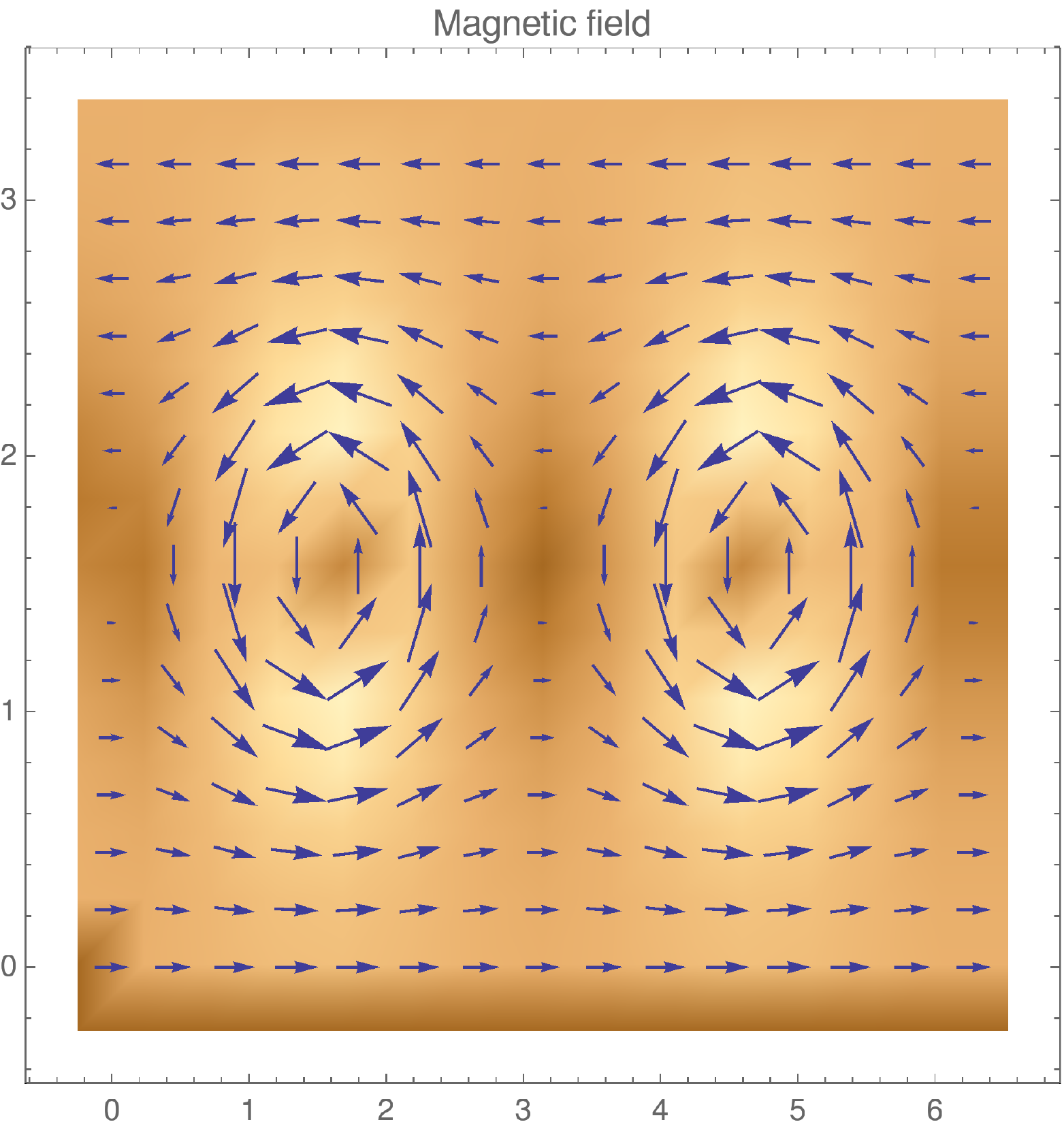}\quad
\includegraphics[scale=0.35]{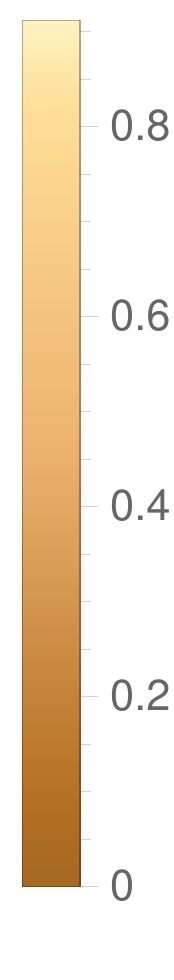}\quad
\includegraphics[scale=0.25]{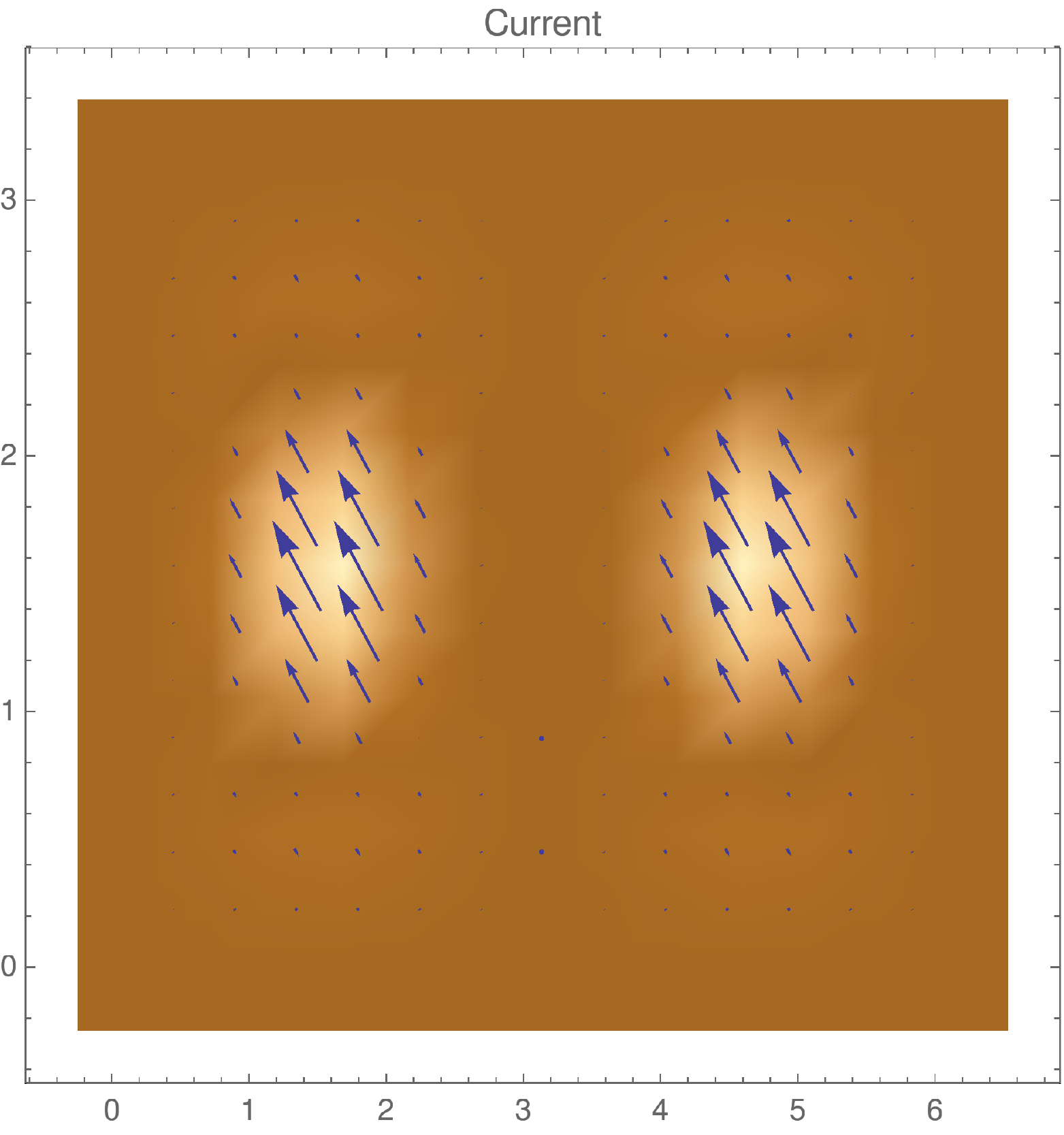}\quad
\includegraphics[scale=0.35]{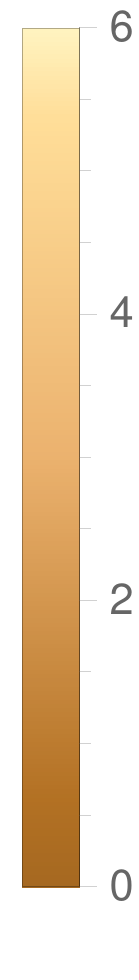} \caption{Electric field, magnetic
field and the current of two gauged solitons, with $n=2$, $m=0$, and $p=q=1$.
The electric and magnetic fields vanish in the center of the tubes while the
current is completely contained inside these.}%
\label{EM}%
\end{figure}

The components of the electric and magnetic fields are given by
\begin{align}
E_{r} &  =-\partial_{r}u\ ,\qquad E_{\theta}=-\partial_{\theta}u\ ,\qquad
E_{\phi}=0\ ,\\
B_{r} &  =\frac{1}{L^{3}}\partial_{\theta}u\ ,\qquad B_{\theta}=-\frac
{1}{L^{3}}\partial_{r}u\ ,\qquad B_{\phi}=0\ .
\end{align}
Figures \ref{Ds} and \ref{EM} show the energy density, the topological
density, the current and the electric and magnetic fields for a simple
configuration of two massless gauged solitons at finite density. The following
boundary conditions have been used
\begin{align}
u(r,0)=u(r,\pi)=0\ , \label{bcu}
\end{align}
while the size of the box has been fixed as $L=1$ and the pions coupling $K=2$
for simplicity. From Figures \ref{Ds} and \ref{EM} one can see that the
present configurations have the shape of three-dimensional tubes (with the
axis along the $\phi$ direction) with a crystalline order: the topological
charge density takes its highest value where the energy density is maximum.
One can also see that the topological charge density vanishes outside the
tubes. The electric and the magnetic fields vanish at the center of the tubes,
and their intensities have maxima on the surfaces of these tubes and they
vanish outside them. On the other hand, the current vanishes outside the tubes.

\begin{figure}[th]
\centering
\includegraphics[scale=0.2]{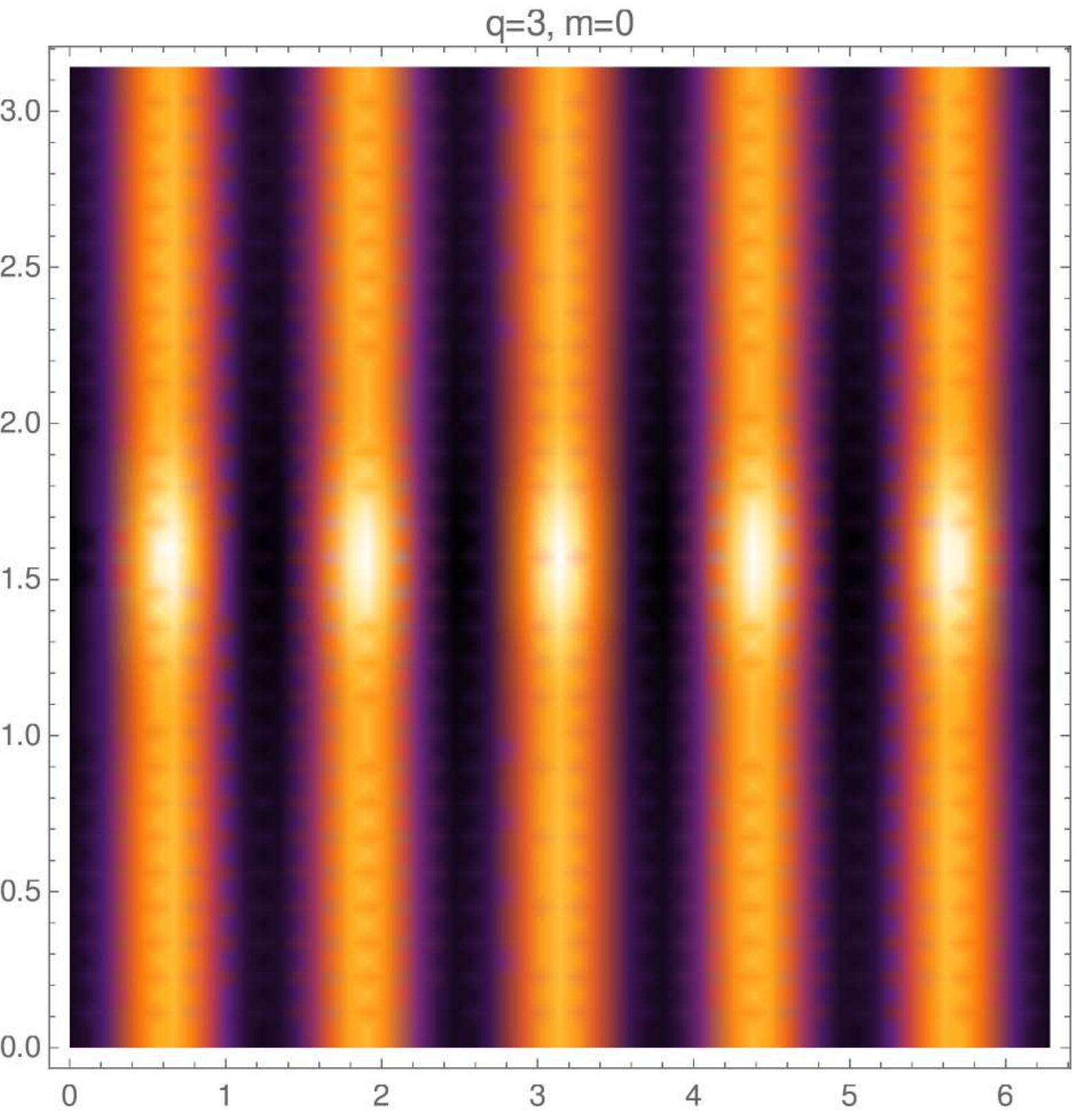}\quad\includegraphics[scale=0.2]{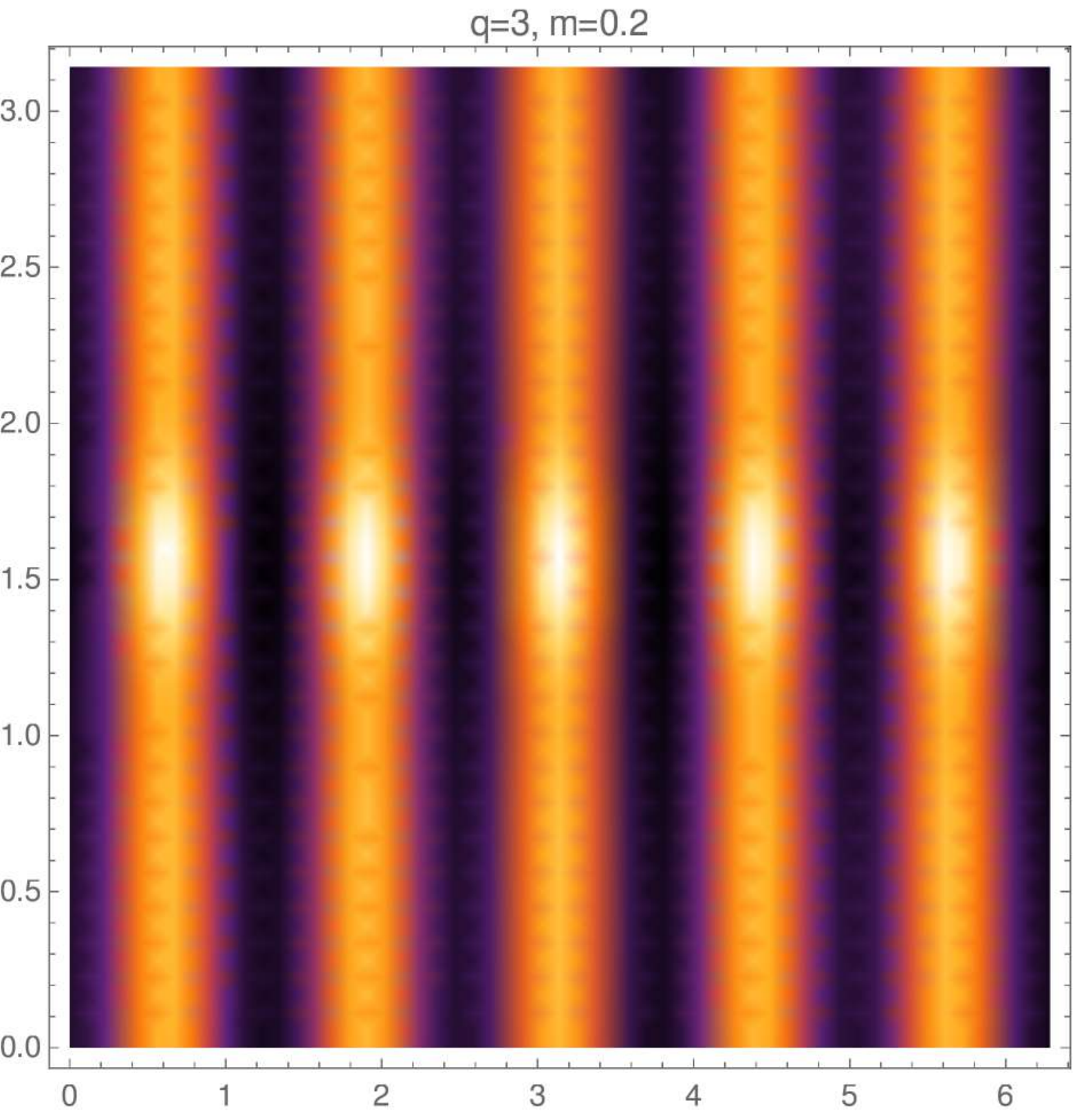}\quad
\includegraphics[scale=0.2]{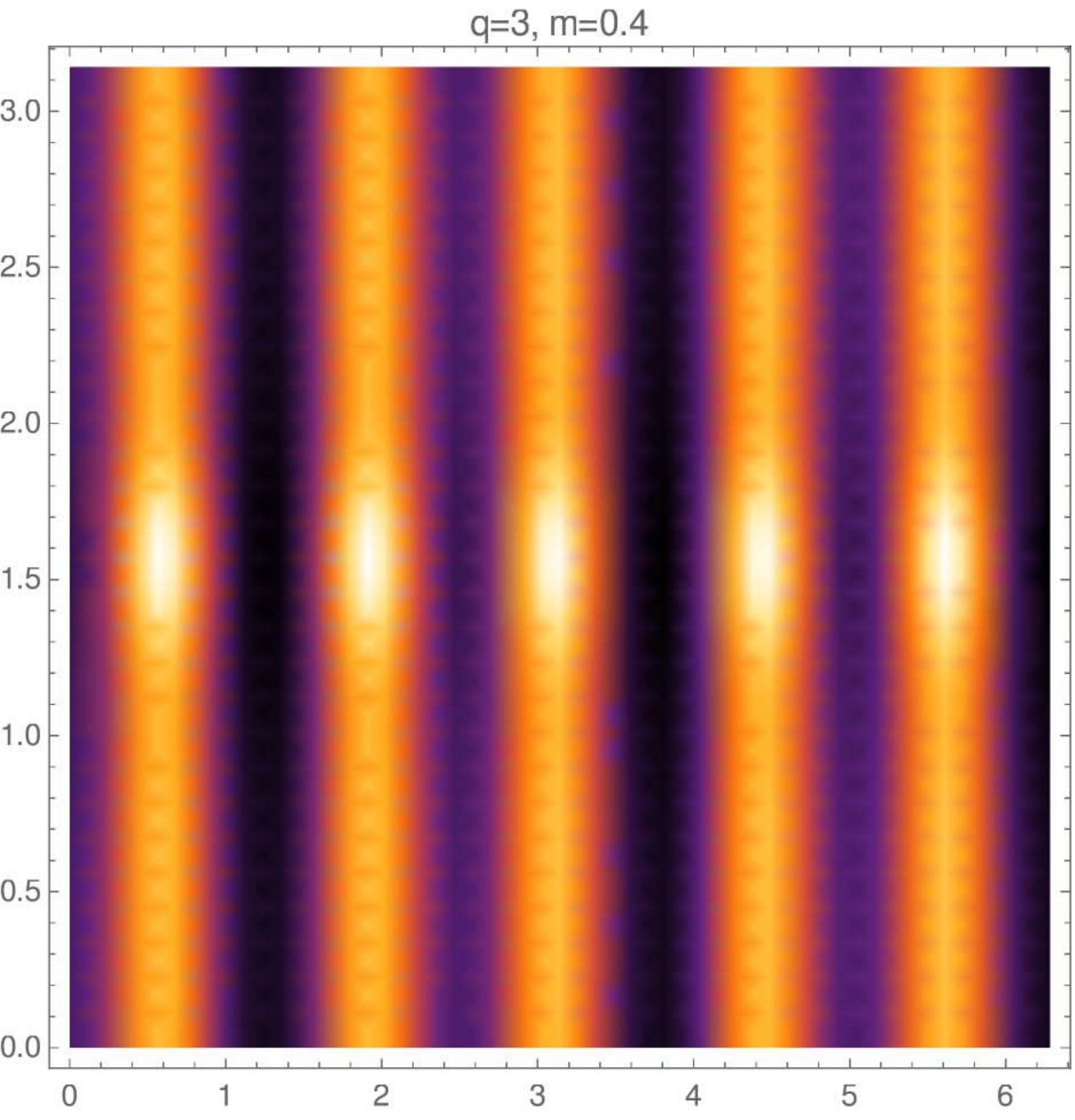}\quad
\includegraphics[scale=0.3]{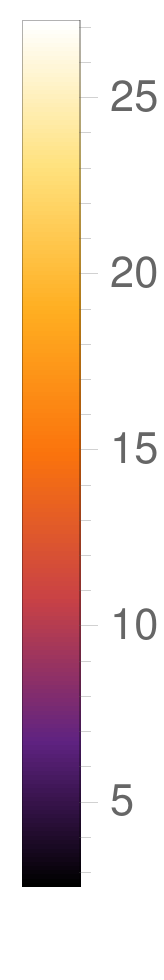}\\%
\includegraphics[scale=0.2]{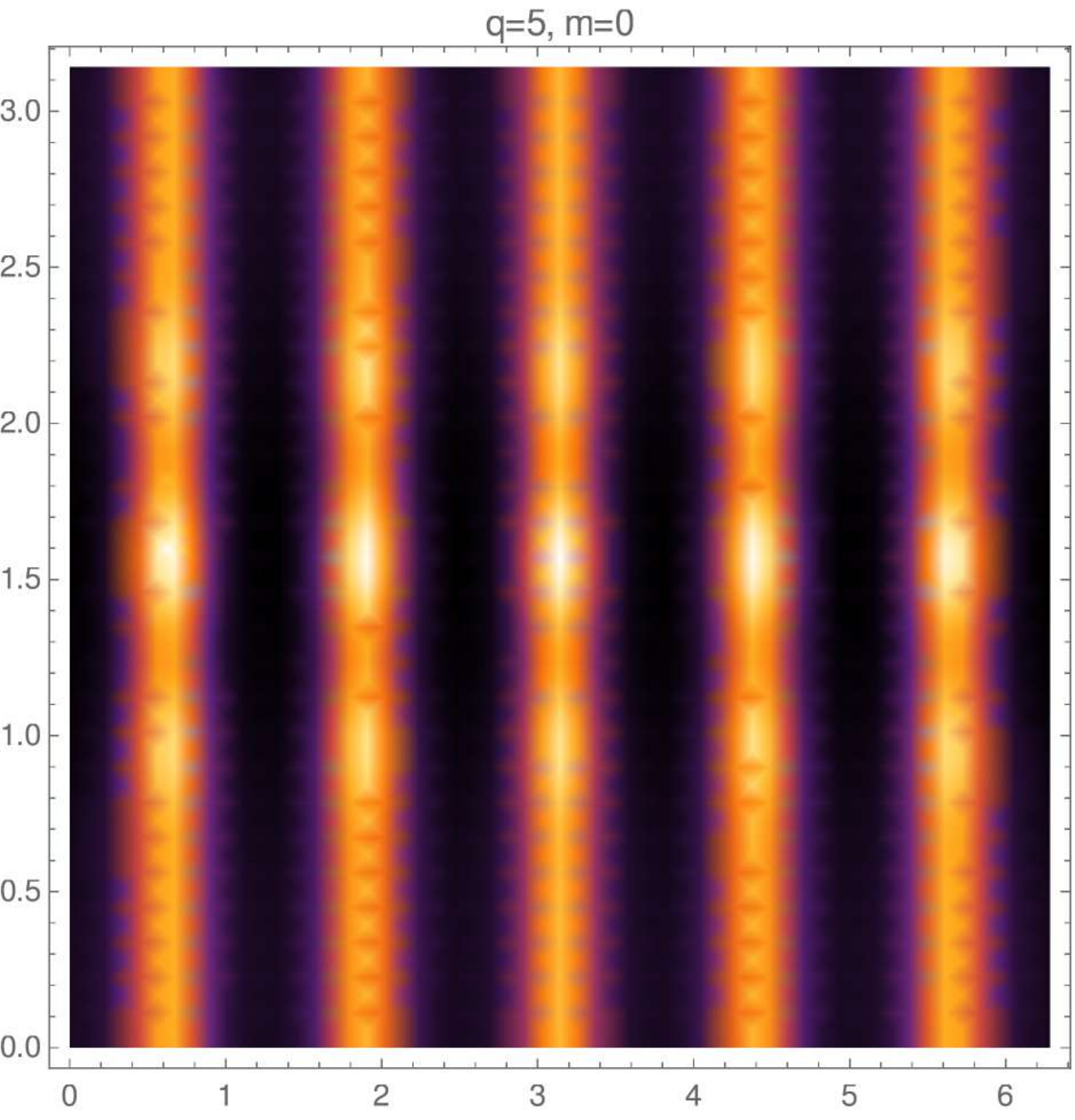}\quad\includegraphics[scale=0.2]{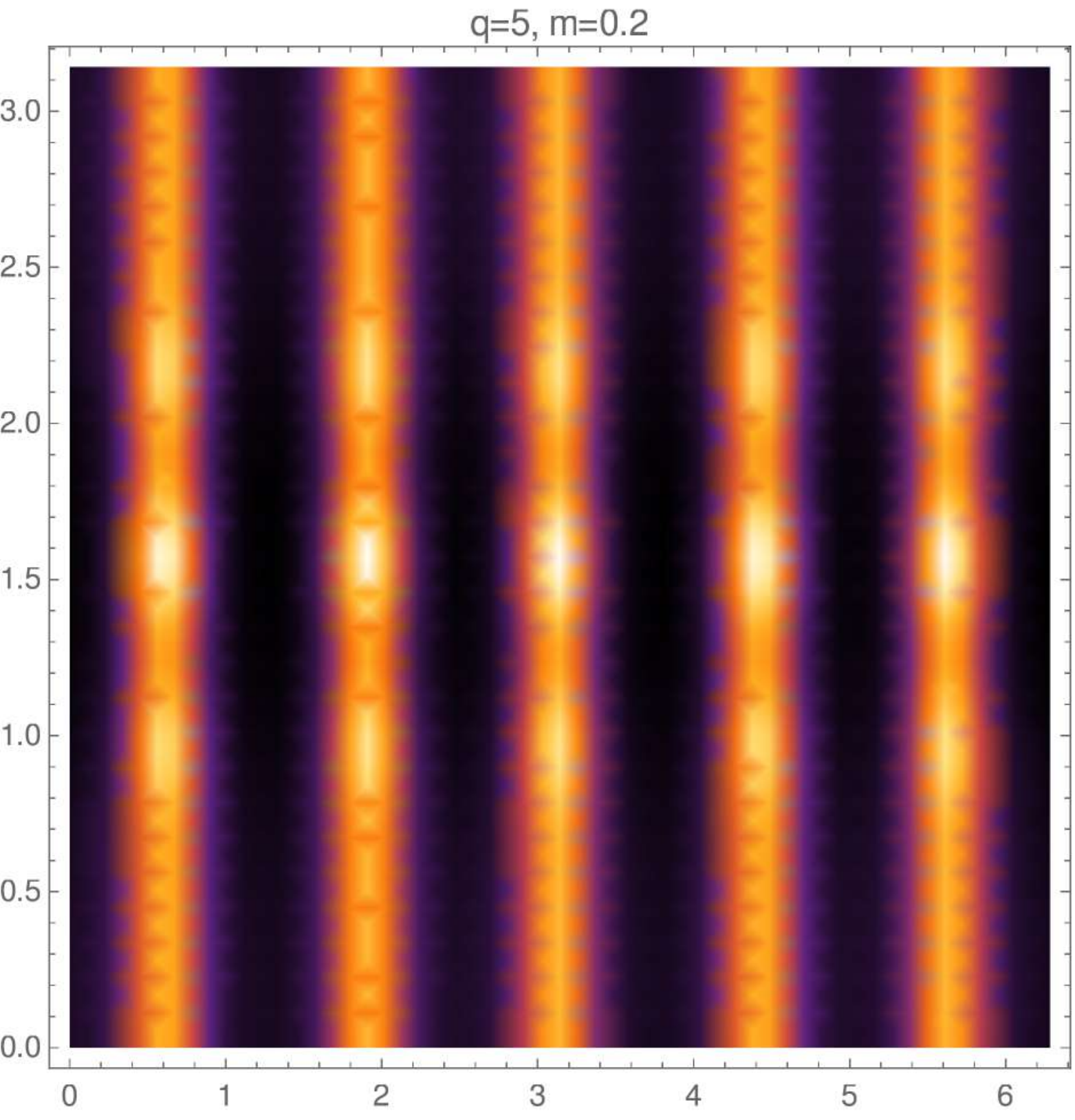}\quad
\includegraphics[scale=0.2]{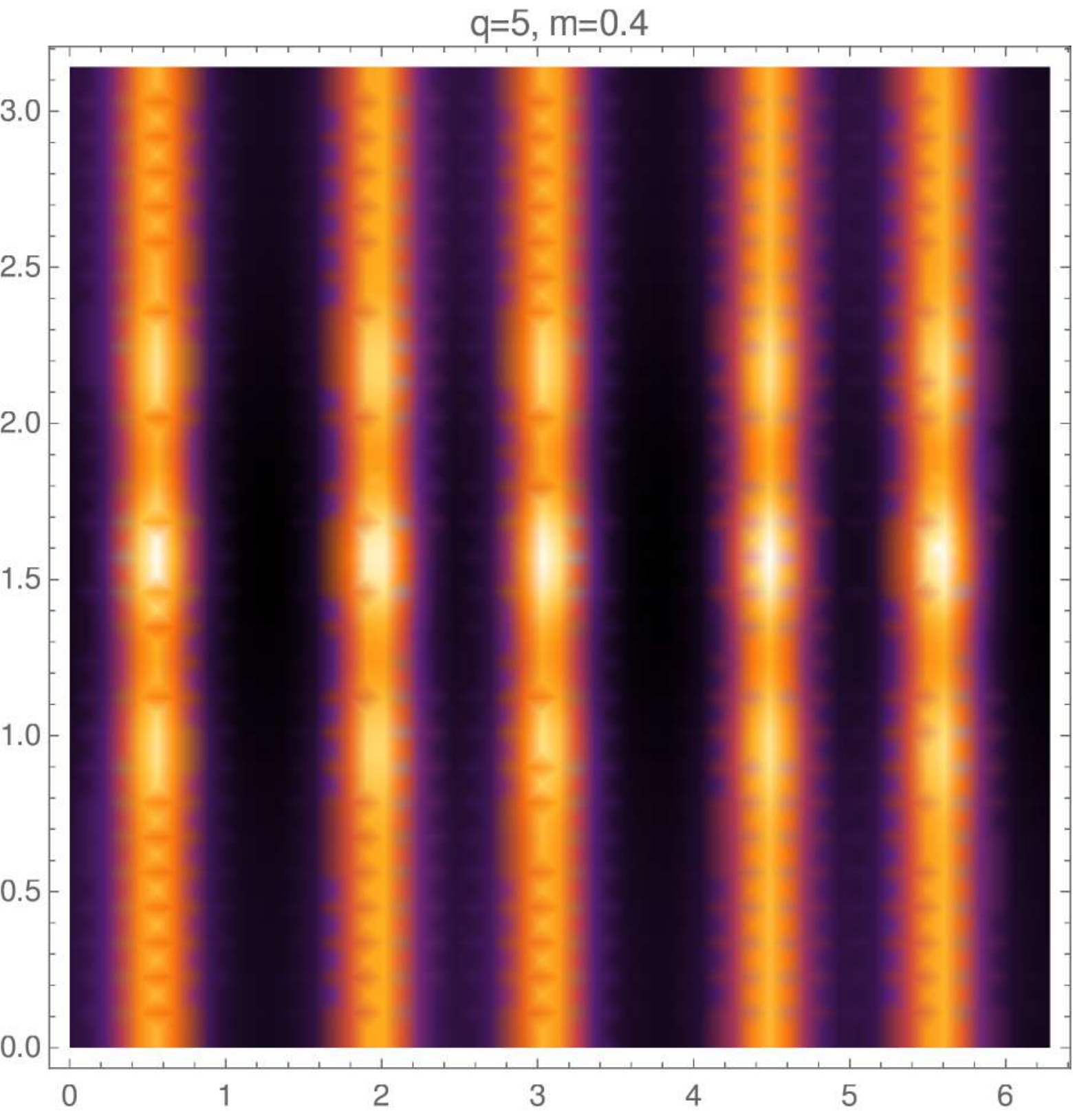}\quad
\includegraphics[scale=0.3]{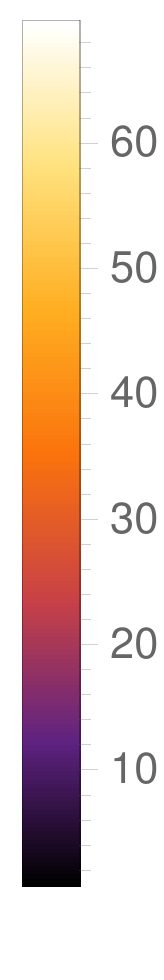}\\%
\includegraphics[scale=0.2]{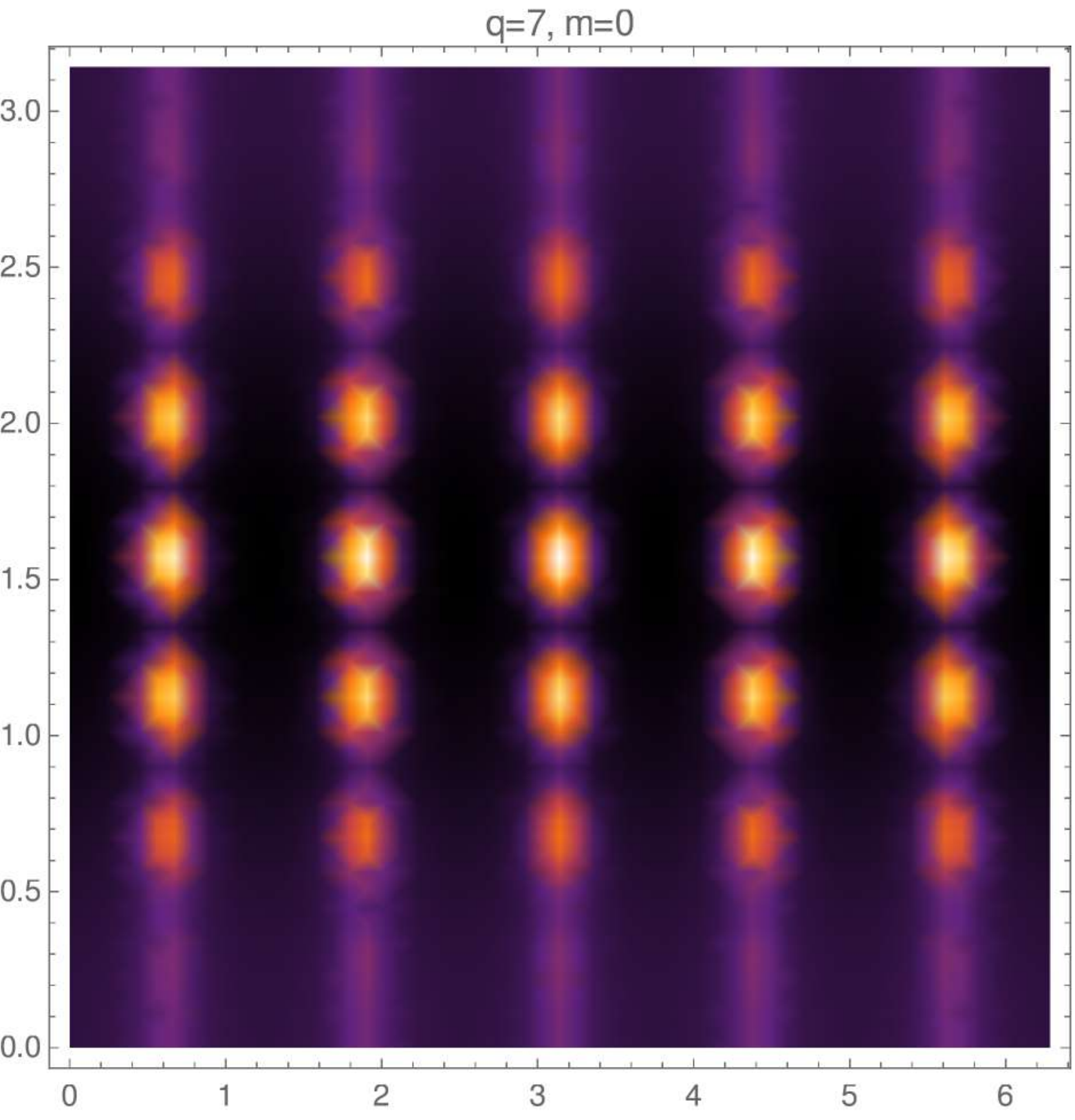}\quad\includegraphics[scale=0.2]{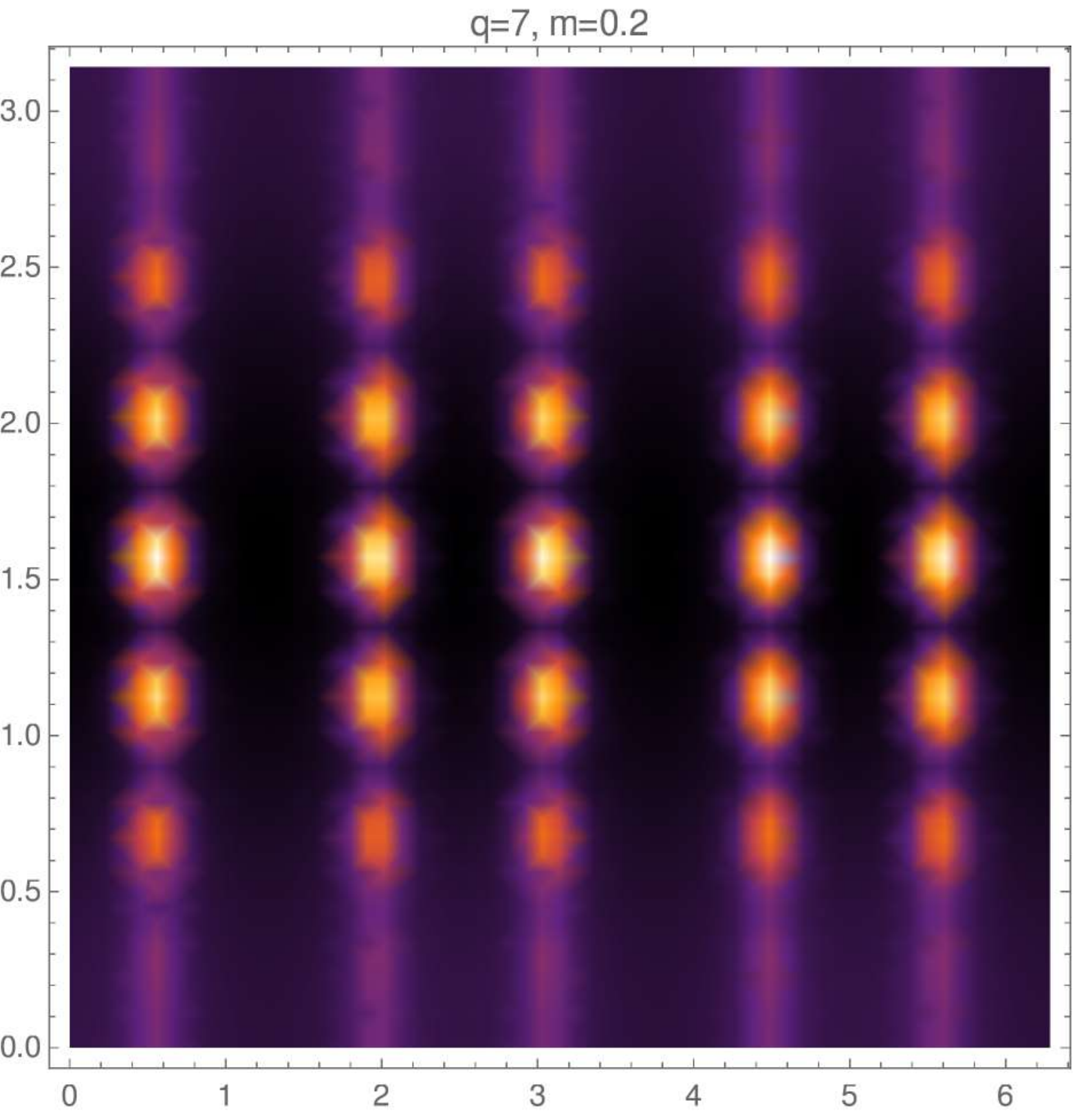}\quad
\includegraphics[scale=0.2]{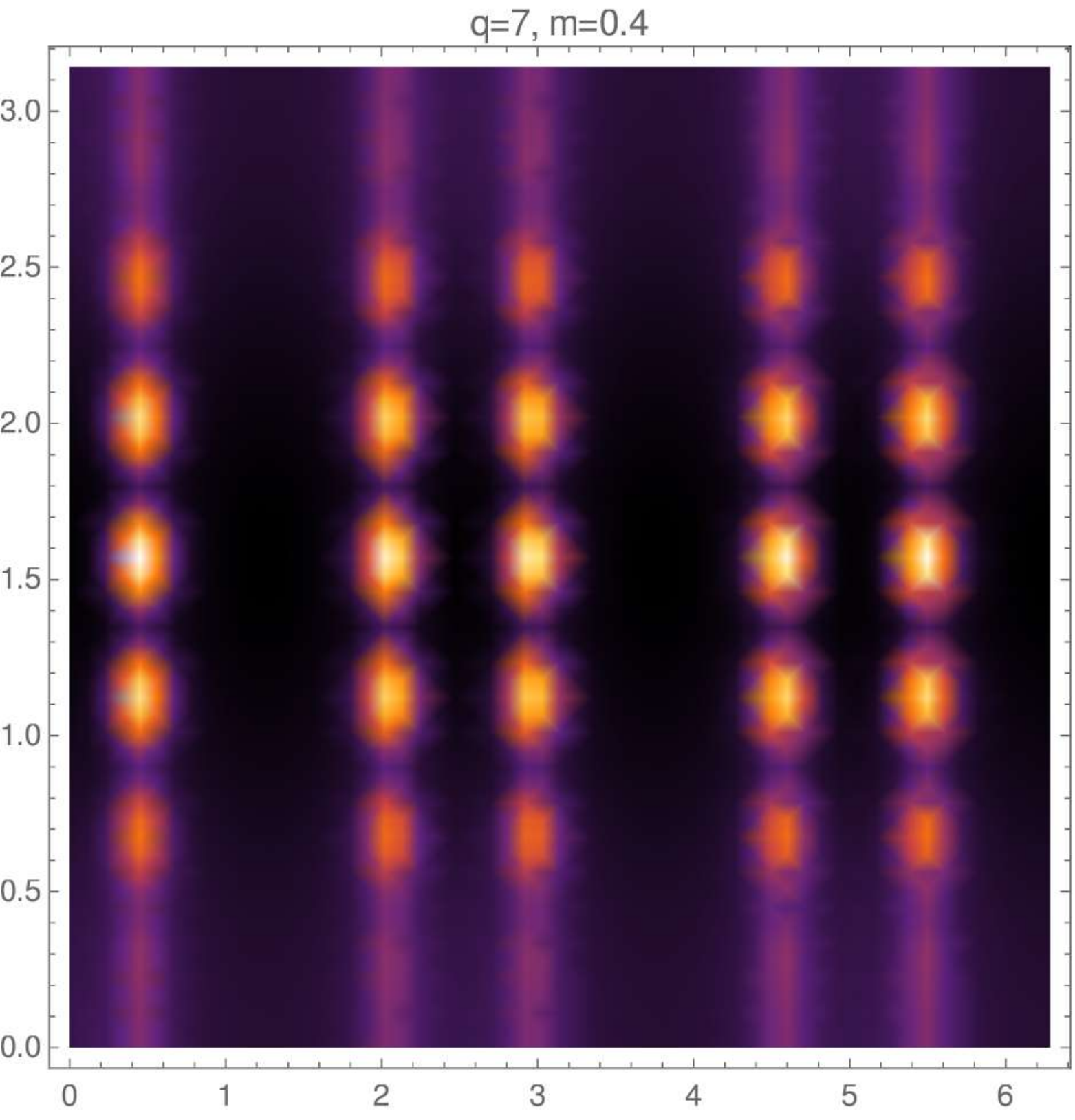}\quad\includegraphics[scale=0.3]{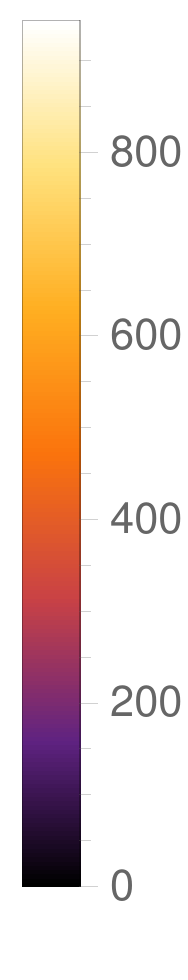}
\caption{Comparison of the energy density for different configurations with
topological charge $B=5$, as function of $q$ and $m$. Here we have considered
$n=5$, $K=2$, $L=1.05$ and $p=1$.}%
\label{Comparison}%
\end{figure}

Figure \ref{Comparison} shows the energy density of different configurations
for a given value of the topological charge ($B=5$ in this case) when the
parameters $q$ and $m$ changes. For this we have imposed again the boundary
conditions in Eq. (\ref{bcu}). One can see that, as the value of $q$
increases, the peaks become more localized and their intensity also increases.
On the other hand, when the value of the mass becomes larger, it is observed
that the spacing between the tubes in the $r$ direction becomes irregular, in
such a way that the tubes are grouped in pairs in this direction.

\subsection{A remark on the stability}

The full stability analysis for the present gauged multi-solitonic
configurations is a quite complicated task even numerically. There are two
main reasons behind this fact: first of all, the field equations linearized
around the gauged solitons constructed here do not reduce to a single master
equation. Secondly, the problem is essentially two dimensional: namely, the
linearized field equations look like a system of two-dimensional coupled
Schrodinger equations which do not allow separation of variables to get
effective one-dimensional problems. However, in the following subsections we
will discuss the stability of these gauged solitons under two particular types
of perturbations.

\subsubsection{Perturbations on the profile}

When the complete set of coupled field equations in a sector with
non-vanishing topological charge reduce to a single equation for the profile
$\alpha$ very sound arguments (see \cite{shifman1}, \cite{shifman2} and
references therein) suggest that a very typical unstable mode corresponds to
perturbations of the profile itself which do not destroy the hedgehog
property:%
\begin{equation}
\alpha\rightarrow\alpha+\varepsilon\Psi\left(  r\right)  \ ,\ \ \ \varepsilon
\ll1\ . \label{pert}%
\end{equation}
One can easily see that if one linearizes Eq. (\ref{sg1}) around a background
solution $\alpha_{0}\left(  r\right)  $ of charge $B=np$ possesses the
following zero-mode: $\Psi\left(  r\right)  =\partial_{r}\alpha_{0}\left(
r\right)  $. If one takes into account Eqs. (\ref{sg2}) and (\ref{sg2.1}), it
turns out that $\Psi(r)$ has no node: therefore $\Psi(r)$ is the perturbation
with the lowest energy. Obviously, this result does not imply that full stability,
nevertheless it is a strong hint as often the stability of solitons with high
topological charge fails precisely due to this type of perturbations.

\subsubsection{Electromagnetic perturbations}

Here we will shortly discuss the behavior of the above class of analytic
gauged solitons under electromagnetic perturbations. As it is well known (see
chapter 4 -in particular, section 4.2- of the classic reference \cite{5b}) at
leading order in the 't Hooft large \textbf{N} expansion, the soliton is
basically unaffected in meson-soliton scattering processes (as it is quite
heavy). Obviously, the same argument applies in the photon-soliton
semiclassical interactions. Consequently, electromagnetic perturbations
perceive the gauged solitons as an effective medium while the perturbations of
the solitons are suppressed by powers of 1/\textbf{N}. In other words, it
makes sense to only consider the reaction of the Maxwell equations to
perturbations around the gauged solitons. So, let's consider the following
electromagnetic perturbation
\[
A_{\mu}^{(0)}\rightarrow A_{\mu}^{(0)}+\varepsilon\bar{A}_{\mu}%
\ ,\ \ \ 0<\varepsilon\ll1\ ,
\]
When one considers general electromagnetic perturbations the contribution at
first order in $\varepsilon$ to the current has the following form
\begin{align*}
J_{\mu}^{\text{pert}} &  =\delta_{\varepsilon}\left\{  \frac{K}{2}%
\text{Tr}\left[  \widehat{O}L_{\mu}\right]  \right\}  =\bar{C}_{\mu}^{\nu}%
\bar{A}_{\nu}\ ,\\
\bar{C}_{\mu}^{\nu} &  =-4K\sin^{2}(\alpha)\sin^{2}(q\theta)\delta_{\mu}^{\nu
}\ ,
\end{align*}
where $\bar{C}_{\mu}^{\nu}$ can be seen as an effective mass-like tensor.
Therefore, the Maxwell equations linearized around the gauged solitons
constructed in the previous sections read%
\begin{equation}
\nabla^{\mu}f_{\mu\nu}=\bar{C}_{\mu}^{\nu}\bar{A}_{\nu}\ ,\ \ f_{\mu\nu
}=\partial_{\mu}\bar{A}_{\nu}-\partial_{\nu}\bar{A}_{\mu}%
\ .\ \label{linmaxwell}%
\end{equation}
In conclusion, the electromagnetic perturbations of the above family of gauged
solitons perceive the background solutions as an effective medium whose
properties are encoded in the mass-like tensor $\bar{C}_{\mu}^{\nu}$ which
explicitly depends on the soliton profile $\alpha$ defined in Eqs.
(\ref{sg1}), (\ref{sg2}) and (\ref{sg2.1}). 

We hope to come back on the analysis of electromagnetic response functions of
this ``effective medium" in a future publication.

\section{Conclusions and perspectives}

The first analytic examples of gauged solitons with non-vanishing topological
charge and with manifest ordered structures in the gauged non-linear sigma
model in (3+1)-dimensions have been constructed. The complete set of seven
coupled non-linear field equations can be reduced (using a judiciously chosen
ansatz) in a self-consistent way to one linear Schrodinger-like equation with
an effective two dimensional periodic potential keeping alive, at the same
time, the baryonic charge. The energy density, the topological charge density
and the current density are periodic and the positions of their peaks manifest
a crystalline order. These configurations describe gauged tubes in which (most
of) the topological charge and total energy are concentrated within
three-dimensional tube-shaped regions whose positions are regular in space.
The electric and magnetic fields vanish at the center of the tubes and take
their maximum values on their surfaces while the electromagnetic current is
contained within these tube-shaped regions. Electromagnetic perturbations of
these gauged tubes perceive them as an effective periodic medium whose
properties can be studied explicitly in term of the solitons profile. The
plots presented here are similar to the ones obtained numerically (see
\cite{pasta1}, \cite{pasta2}, \cite{pasta3}, \cite{pasta4} and references
therein) in the description of nuclear spaghetti. On the other hand, in the
present case the gauged solitons (besides to be analytic) also include the
effects of their own electromagnetic field in a self-consistent way. These
results open many interesting possibilities. One can, for instance, analyze
the electromagnetic response functions of these gauged solitons in the
approximation in which they are perceived as an effective medium by
electromagnetic perturbations. It is also very interesting to analyze the
semi-classical quantization of these gauged tubes following the classic
references \cite{witten0}, \cite{bala0}, \cite{Bala1}, \cite{ANW} (however
some of the techniques in the above references cannot be applied directly here
since the present gauged solitons are not static, although the corresponding
energy-momentum tensor is). We hope to come back on these very nice topics in
future publications.

\subsection*{Acknowledgements}

This work has been funded by Fondecyt grant 1160137. A.V. appreciates the
support of CONICYT Fellowship 21151067. This work is partially supported by
the National Research Foundation of Korea funded by the Ministry of Education
(Grant 2018-R1D1A1B0-7048945). The Centro de Estudios Cient\'{\i}ficos (CECs)
is funded by the Chilean Government through the Centers of Excellence Base
Financing Program of Conicyt.

\section*{Appendix: Obtaining the field equations}

Explicitly, the $SU(2)$ scalar field, according to our ansatz defined in
Eqs. (\ref{standard1}), (\ref{standard1.1}), (\ref{ans1}) and
(\ref{gaugcrystal2}), is given by
\begin{align*}
U=
\begin{pmatrix}
\cos(\alpha)+i\cos(q\theta)\sin(\alpha) & ie^{-\frac{ip}{L}(t-L\phi)}%
\sin(q\theta)\sin(\alpha)\\
ie^{\frac{ip}{L}(t-L\phi)}\sin(q\theta)\sin(\alpha) & \cos(\alpha
)-i\cos(q\theta)\sin(\alpha)
\end{pmatrix}
.
\end{align*}
It follows that the components of $L_{\mu}=U^{-1}D_{\mu}U$ read%
\begin{align*}
L_{t}  &  =\frac{P}{L}
\begin{pmatrix}
i\sin^{2}(q\theta) \sin^{2}(\alpha) & E^{(+)}F^{(-)}\\
-E^{(-)}F^{(+)} & -i\sin^{2}(q\theta)\sin^{2}(\alpha)
\end{pmatrix}
\ , \quad L_{r} =i\alpha^{\prime}%
\begin{pmatrix}
\cos(q\theta) & E^{(+)}\sin(q\theta)\\
E^{(-)}\sin(q\theta) & -\cos(q\theta)
\end{pmatrix}
\ ,\\
L_{\theta}  &  =q\sin(\alpha)
\begin{pmatrix}
-i\sin(q\theta)\cos(\alpha) & E^{(+)}G^{(+)}\\
E^{(-)}G^{(-)} & i\sin(q\theta)\cos(\alpha)
\end{pmatrix}
\ , \quad L_{\phi}= P
\begin{pmatrix}
-i\sin^{2}(q\theta)\sin^{2}(\alpha) & -E^{(+)}F^{(-)}\\
E^{(-)}F^{(+)} & i\sin^{2}(q\theta)\sin^{2}(\alpha)
\end{pmatrix}
\ ,
\end{align*}
where we have defined
\begin{align*}
P  &  = p-2Lu \ , \qquad F^{(\pm)} = \biggl(\cot(\alpha) \pm i\cos(q\theta)
\biggl)\sin(q\theta)\sin^{2}(\alpha) \ ,\\
E^{(\pm)}  &  = e^{\pm i\frac{p}{L}(L\phi-t)} \ , \qquad G^{(\pm)}%
=i\cos(q\theta)\cos(\alpha)\pm\sin(\alpha) \ .
\end{align*}

On the other hand, varying the Lagrangian w.r.t the $U$ field we obtain
\begin{align*}
\delta\mathcal{L} &  =\text{Tr}\biggl[\frac{K}{4}\delta(L^{\mu}L_{\mu}%
)-m^{2}\delta(U+U^{-1})\biggl]\\
&  =\text{Tr}\biggl[\frac{K}{2}L^{\mu}\delta L_{\mu}-m^{2}(\delta U+\delta
U^{-1})\biggl]\ .
\end{align*}
For the non-linear sigma model term, we use
\begin{align*}
\delta(UU^{-1}) &  =0\quad\rightarrow\quad\delta U^{-1}=-U^{-1}\delta
UU^{-1}\ ,\\
D_{\mu}(UU^{-1}) &  =0\quad\rightarrow\quad D_{\mu}U^{-1}=-U^{-1}D_{\mu
}UU^{-1}=-L_{\mu}U^{-1}\ ,
\end{align*}
so that
\begin{align}
\delta L_{\mu} &  =\delta(U^{-1}D_{\mu}U)\nonumber\\
&  =-U^{-1}\delta UU^{-1}D_{\mu}U+U^{-1}\delta D_{\mu}U\nonumber\\
&  =-U^{-1}\delta UL_{\mu}+L_{\mu}U^{-1}\delta U+D_{\mu}(U^{-1}\delta U)\ .
\end{align}
From the above, the variation of the non-linear sigma model term (using the
cyclicity of the trace for the first two terms) becomes
\begin{align}
\text{Tr}(L^{\mu}\delta L_{\mu}) &  =\text{Tr}(L^{\mu}D_{\mu}(U^{-1}\delta
U))\nonumber\\
&  =-\text{Tr}(D_{\mu}L^{\mu}U^{-1}\delta U)+\text{Tr}(D_{\mu}(U^{-1}\delta
UL^{\mu}))\ ,\label{var1}%
\end{align}
where we have integrated by parts. Now, introducing this in the variation of
the Lagrangian, we obtain
\begin{align*}
\delta\mathcal{L} &  =\text{Tr}\biggl[-\frac{K}{2}D_{\mu}L^{\mu}U^{-1}\delta
U-m^{2}(\delta U-U^{-1}\delta UU^{-1})\biggl]\quad+\quad\text{Boundary term}\\
&  =-\text{Tr}\biggl[\frac{K}{2}D_{\mu}L^{\mu}U^{-1}\delta U+m^{2}%
(UU^{-1}\delta U-U^{-1}U^{-1}\delta U)\biggl]\quad+\quad\text{Boundary term}\\
&  =-\text{Tr}\biggl[\left(  \frac{K}{2}D_{\mu}L^{\mu}+m^{2}(U-U^{-1})\right)
U^{-1}\delta U\biggl]\quad+\quad\text{Boundary term}\ ,
\end{align*}
and because $U\neq0$ and the first factor in the trace is in the algebra, it
is necessarily that
\[
D_{\mu}L^{\mu}+\frac{2m^{2}}{K}\left(  U-U^{-1}\right)  =0\ .
\]
Replacing the $L_{\mu}$ in the previous equation we lead to the equation in
Eq. (\ref{sg1}).

\end{document}